\documentclass[aps,psfig,epsfig,twocolumn]{revtex4}
\usepackage{epsfig}

\newcommand{\tu}{\widetilde}

\newcommand{\pa}{\partial}

\begin{document}
\draft
\title{
\vspace{1cm} Quantum Horizons   }
\author{Hua Bai$^2$\footnote{E-mail address: huabai@mail.ustc.edu.cn},
Mu-Lin Yan$^{1,2}$\footnote{E-mail address:
mlyan@staff.ustc.edu.cn (corresponding author)}}
\address{
$^1$CCAST (World Lab), P.O. Box 8730, Beijing, 100080,  China \\
$^2$Interdisciplinary Center for Theoretical Study, University of
Science and Technology of China, \\ Hefei, Anhui 230026, China}

\begin{abstract}
 Treating macro-black hole as quantum states, and using Brown-York
 quaselocal gravitational energy definition
and Heisenberg uncertainty principle, we find out the classical
horizon with singularity spreads into a quantum horizon in which
the space-time is non-commutative and the spread range is
determined dynamically. A Quantum Field Theory (QFT) model in
curved space with quantum horizon is constructed. By using it, the
black hole entropy and the Hawking temperature are calculated
successfully. The $\phi-$field mode number is predicted and our
quantum horizon model favors to support the Minimal
Super-symmetric Standard Model.
\end{abstract}
\maketitle
  Physicists are intrigued by the idea  that if one
attempts to quantize the gravitational force, one should also
consider the question how Quantum Mechanics affects the behavior
of black hole\cite{thooft04}. The Quantum Mechanics(QM) and the
Quantum Field Theory (QFT) in the curved space-time with
(classical) event horizon provides a framework to understand the
statistic mechanics origin of the thermodynamics of non-extreme
black hole, and serves as a powerful tool to calculate its entropy
and the Hawking radiation
temperature\cite{thooft04}\cite{wald}\cite{Hawking74}\cite{Ruffini}
\cite{thooft85}\cite{thooft96}. However this theory is seriously
plagued by the singularity at the horizon: Firstly, in the entropy
calculations, an ultraviolet cutoff (or brick wall) has to be put
in by hand\cite{thooft85} (even though it could be done in a
proper way \cite{thooft04}). Otherwise, the entropy will be
divergent; Secondly, in the Hawking radiation derivations, one
should also artificially use an analytic continuing trick to go
over the singularity associating with that
horizon\cite{Hawking74}\cite{Ruffini}\cite{Wilczek}. This disease
spoils our understanding of Hawking radiation as quantum
tunnelling effects. Thus we face an unsatisfactory situation that
if once set up a brick wall for getting the entropy rightly, we
will have no way to do analytic continuing and hence no way to
derive the Hawking temperature. On the other hand, if once
withdrew the wall (or the ultraviolet cutoff), we will have no way
to handle the entropy even though the Hawking temperature deriving
becomes doable. Consequently, a self-consistent QFT (or QM) model
in curved space to calculate both the black hole's entropy and the
Hawking temperature simultaneously and rightly is still in
absence.

  Recently, we made some remarks\cite{by03} on 't Hooft's brick wall
model which is one of QFTs in curved space with (classical)
horizon, and provided a heuristic discussion for how to go over
the obstacle mentioned in above. The point of our idea is to
consider a semi-classical effect on the horizon , which naturally
led to the space-time non-commutativity on the horizon, and
dramatically and successfully led to solve the puzzles
 due to the singularity at the classical horizon. We
will call such kind of event horizon on which the space-time is
non-commutative as {\it Quantum Horizon} in this letter.  As a
consequence in ref.\cite{by03}, we simultaneously got the
Schwarzschild black hole's Hawking temperature and entropy in a
probing QFT-model with a quantum horizon. The results are
qualitatively right. In this treatment, the ordinary classical
horizon spreads, and hence the troubles  due to the singularity on
the horizon  disappear. This seems to be a strong sign to indicate
that the quantum effects on the horizon are important for the the
thermodynamics of black hole. In other wards, the quantum horizon
conception may play a essential role for solving the puzzle
brought by the classical horizon singularity. In this letter we
will pursue this idea more accurately, and construct general
QFT-model for stationary black hole in curved space with quantum
horizon, and some Planck scale physics will be discussed.

In the refs.(\cite{thooft04})(\cite{thooft96}), 't Hooft shows
that assuming the process of the Hawking particle emissions is
governed by a Schr\"{o}dinger equation, and hence describing the
black hole as wave function $|M\rangle_{BH}$, then the first law
of thermodynamics of black hole can successfully be derived out,
and the  Bekenstein-Hawking expression  of the black hole entropy
can be achieved:
\begin{equation}\label{s1}
S={A \over 4G},
\end{equation}
where $A=4\pi r_H^2$ is the horizon area (e.g., see
eqs.(3.1)-(3.9) in ref.(\cite{thooft04})). These studies indicate
the black hole can be treated as a quantum state with high
degeneracy instead of a classical object.
 Following this point of view, and according to quantum mechanics principle, then we
should conclude its energy and its corresponding conjugate time
$t$ can not be simultaneously measured exactly (Heisenberg
uncertainty principle). Namely, when treating $E$ and $t$ as
operators, we have $[t, E]=i$.  In other hand, it has been
generally believed that the states are situated on the horizon,
and outside it the densities of the states become tiny fast.
Therefore, the energy of states $E$ should approximatively be the
hole's gravitational energies measured at the location of the
horizon instead of infinity (i.e., instead of the ADM-mass of the
hole\cite{by2}). It has long been aware the gravitational energies
are quasilocal in the general relativity. In this letter, we
follow Brown-York's quasilocal energy(QLE)
defination\cite{Brown93}, and the energy of the black hole
$E_{QLE}$ is coordinate-dependent. Then, the states energy $E$ is
$E=E_{QLE}(r=r_H)$, where $r_H$ is the radio of black hole (or the
radio of event horizon), and it is a function of the parameters of
the hole.

  In this letter, we study the
large and non-extreme $Reissner-Nordstr\ddot{o}m$ black hole. Its
metric generically can be expressed as follows
\begin{equation}\label{metric}
ds^2=-B(r)dt^2+B(r)^{-1}dr^2+r^2(d\theta^2+\sin^2\theta
d\varphi^2),
\end{equation}
where $B(r)=(r-r_-)(r-r_H)/r^2$, . The definitions of the $r_H$
and $r_-$ are
\begin{eqnarray}
r_\pm=G(M\pm\sqrt{M^2-Q^2}),\;\;\;\;\;\;\;\;\; r_H=r_+.
\end{eqnarray}
where $M$ and $Q$ are the mass and charge of the black hole. The
large and non-extreme block hole conditions are $r_H\gg l_p$, and
$r_H-r_-\gg l_p$, where $l_p=\sqrt{G}$ is the Planck length(in
this letter $\hbar=c=1$) . The time Killing vector, $\pa t$, is
everywhere. Following Brown-York's quasilocal energy ($E_{QLE}$)
definition \cite{Brown93}, the energy of the black hole reads
\begin{equation}
E\left( r\right) _{QLE}=r\left(1-\sqrt{B(r)}\right)/G.
\end{equation}
 And as $r=r_H$, we have $B=0$ and
$E(r_H)_{QLE}=r_H/G$, we get $E=E_{QLE}(r=r_H)={r_H/G}$. With this
equation, we must conclude that the uncertainty of $E$ implies
that of $r_H$, and then we have
\begin{equation}\label{NC}
[t,r]|_{r=r_H}=il_p^2.
\end{equation}
 This equation implies
that the radial coordinate is noncommutative with the time at the
horizon. The corresponding uncertainty relation for them is
$(\Delta t)(\Delta r)|_{r \sim r_H} \sim l_p^2$. In other hands,
due to quantum measurement effects, $r_H$ spreads into a range of
$\{r_H-\Delta , r_H+\Delta \}$, here for simplicity we denote
$\Delta r=\Delta$  and this spread range will be denoted as
$\mathcal{R}_I$, and ones outside $\mathcal{R}_I$ are
$\mathcal{R}_{II}$. Thus, the previous classical horizon spreads
into the quantum horizon, and Eq.(\ref{NC}) is extended as
$[t,r]|_{r\in \mathcal{R}_I}=il_p^2$. In the range
$\mathcal{R}_I$, the QFT is noncommutative, and in the space-time
besides that range QFT back to be commutative as usual since
$[t,r]|_{r\in \mathcal{R}_{II}}=0$.

According to our aforementioned discussion, we construct a model
with noncommutative $\phi$-field in  $\mathcal{R}_I$ and with
commutative $\phi$-fields in  $\mathcal{R}_{II}$.  It is essential
that the boundary between $\mathcal{R}_{I}$ and $\mathcal{R}_{II}$
denoted by $\Delta$ should be an intrinsic quantity of the model,
and should be determined by dynamics of the model. The
$\phi$-field equation of motion in noncommutative background is
different from the ordinary Klein-Gorden equation in curved space.
Comparing the equation of motion in $\mathcal{R}_I$ with the
later, we get an effect metric $\tu{g}_{\mu\nu}$. Remarkably, it
will be shown below that $\widetilde{g}_{tt}$ has two new
singularities , i.e.,in which $\widetilde{g}_{tt}(r_H\pm
\Delta_0(E_\phi))=0$. This means that to the fields with energy
$E_\phi$, $\phi(E_\phi,r)$, its red-shifting on the $(r_H\pm
\Delta _0(E_\phi))$ surface is infinite. As discussed in
\cite{by03}, these two red-shift surfaces are  the boundaries
between the $\mathcal{R}_I$ and $\mathcal{R}_{II}$, and hence we
have $\Delta=\Delta_0(E_\phi)$.

With the above, we now consider the field as a probe in the region
$\mathcal{R}_I $ but moving in a non-commutative background. We
rewrite space-times coordinates commutative relation as follows
\begin{eqnarray}\label{NCL}
[x^i,x^j]=i\Theta\varepsilon^{ij},&& (i,j=0,1),\;(x^0=t,x^1=r),
  \\ \nonumber
[x^k,x^\mu]&=&0, \    \ (k=2,3;\mu=0,1,2,3)
\end{eqnarray}
where $\Theta={l_p^2}$ and $\varepsilon^{ij}$ is antisymmetrical
with $\varepsilon^{01}=1$. The star product of two function $f(x)$
and $g(x)$ is given by the Moyal
formula\cite{Witten}\cite{Seiberg}:
\begin{equation}\label{sp1}
(f\star g)(x)= \exp\left[{i\over
2}\Theta\varepsilon^{ij}\frac{\partial}{\partial
x^i}\frac{\partial}{\partial y^j}\right]f(x)g(y)|_{y = x}.
\end{equation}
In terms of above notations, the action of the QFT-model in curved
space with {\it quantum horizon} is set to be as follows:
\begin{eqnarray}\label{Nac0}
I=I_{\mathcal{R}_I}+I_{\mathcal{R}_{II}}
\end{eqnarray}
where
\begin{eqnarray}\label{Nac}
I_{\mathcal{R}_I}  = -{1\over 2}\int  d^4 x \sqrt{-g^\star}  \star
g^{\mu\nu}\star (\pa_{\mu}\phi
\star \pa_{\nu}\phi),\;\hskip-0.05in &r&\hskip-0.05in\in\mathcal{R}_I\\
\label{Nac1} I_{\mathcal{R}_{II}}=-{1\over 2}\int d^4 x \sqrt{-g}
g^{\mu\nu} (\pa_{\mu}\phi \pa_{\nu}\phi),\;r\hskip-0.05in &\in&
\hskip-0.05in \mathcal{R}_{II}
\end{eqnarray}
where $g^\star$ means that the determinate of the metric
calculated by using the star product.To the case of $[t, r]\neq
0$, we have the relations as follows:
\begin{equation}\label{r1}
f(r)\star h(r,t)=f(r)h(r,t)+\pa_t F(t,r)
\end{equation}
\begin{equation}\label{r2}
f_1(r)\star f_2(r)\star\cdots\star f_n(r)\star
h(r,t)=\left[f_1(r)f_2(r)\cdots f_n(r)\right] \star h(r,t)
\end{equation}
where the $f_n$ is the function only about the $r$, and $\pa_t
F(t,r)$ are some total $t-$derivative terms, when which emerges in
an action and will have no contributions to the equation of
motion, and hence can be dropped in the action, i.e., $\int d^4x
\pa_t( \cdots)\Rightarrow 0$.

To Reissner-Nordstr$\ddot{\rm o}$m black hole, the metric is given
by eq.(\ref{metric}). Because the metric $g_{\mu\nu}$ is not the
function of $t$ and noting $[r,\varphi]=[r,\theta]=[\varphi,
\theta]=0$ (see eq.(\ref{NCL})), and by eq.(\ref{r2}) we have
\begin{equation}\label{r3}
\sqrt{-g^\star}\star g^{\mu\nu}=\sqrt{-g}g^{\mu\nu}.
\end{equation}
 Then, by (\ref{r3}) and (\ref{r1}), and considering $\phi=\phi(t,r,\theta,\varphi)$,
 the action of (\ref{Nac}) is reduced to be
\begin{equation}\label{action1}
I_{\mathcal{R}_I}=-{1\over 2}\int d^4x
\sqrt{-g}g^{\mu\nu}(\partial_\mu\phi\star
\partial_\nu\phi)\;\;\;r\in\mathcal{R}_I.
\end{equation}
  By using
eq.(\ref{sp1}), we evaluate the start product in the action of
(\ref{action1})
   and cast the action in the ordinary product. By
this, the noncommutative effect can be absorbed into an equivalent
background metric. In other words, we first take the
semi-classical quantum effect into consideration. This effect is
then realized through the non-commutative geometry. Finally, this
effect is further through an effective background but in an
ordinary geometry.

Using the $\delta I/\delta \phi =0$, the equation of motions can
be derived. In the range $\mathcal{R}_{II}$ (i.e., $r\in
\mathcal{R}_{II}$),  $\delta I/\delta \phi =\delta
I_{\mathcal{R}_{II}}/\delta \phi =0$, then the equation of motions
is as usual Klein-Gorden equation in curved space, i.e.,
\begin{equation}\label{EMII}
\pa_\mu(\sqrt {-g}g^{\mu\nu}\pa_\nu\phi)=0.
\end{equation}
 In the range $\mathcal{R}_{I}$ (i.e., $r\in
\mathcal{R}_{I}$), however, the equation of $\delta I/\delta \phi
=\delta I_{\mathcal{R}_{I}}/\delta \phi =0$ will be much longer
than eq.(\ref{EMII}). The calculations are straightforward, and
the result is as follows
\begin{eqnarray}\label{NMT1}
&&\pa_\mu(\sqrt {-g}g^{\mu\nu}\pa_\nu\phi)  \nonumber \\
&+& \pa_r\left[{1\over 2!}\left({i\Theta\over 2}\right)^{2}(\sqrt
{-g}g^{rr})_{,rr}{{\pa_t^2}}\pa_r\phi \right]\\
\nonumber &+&\pa_t\left[\sum_{n=1}^\infty{1\over
2n!}\left({i\Theta\over 2}\right)^{2n}(\sqrt
{-g}g^{tt}),_{\underbrace{r\cdots
r}\limits_{2n}}\underbrace{{\pa_t\cdots
\pa_t}}\limits_{2n}\pa_t\phi \right]
 =0
\end{eqnarray}
where $(\sqrt{ -g}g^{tt}),_{\underbrace{r\cdots r}\limits_{n}}$
stands for $\underbrace{{\pa_r\cdots \pa_r}}\limits_{n}(\sqrt{
-g}g^{tt})$, etc. Note all terms in the Moyal product-expansion
have been included, and hence the equation is strict.
 For a given energy mode, i.e., assuming the WKB
approximation wave function $\phi(t,r,\theta,\varphi)=e^{-itE_\phi
-i\int k(r) dr}Y_{lm}(\theta,\varphi)$, the effective metric can
be  from the following equation of motion for the scalar field
once the star product is evaluated:
\begin{eqnarray}\label{em0}\nonumber
&&-\{{r^4\sin \theta \over (r-r_-)(r-r_H)} +{\sin
 \theta \over r_H-r_-}  [\tu{\Delta}(E_\phi)^{2} (r_H-r_-) \\\nonumber && + {r_H^4 \over r-r_H} \left({1\over
1-\left({\tu{\Delta}(E_\phi) \over r-r_H} \right)^2}-1\right) \\
\nonumber && - {r_-^4 \over r-r_-} \left({1\over
1-\left({\tu{\Delta}(E_\phi) \over r-r_-} \right)^2}-1\right) ]\}
\phi,_{tt}\\ \nonumber &&+\sin\theta
[(r-r_-)(r-r_H)+\tu{\Delta}(E_\phi)^2]\phi ,_{rr} \\ && +
\pa_\theta (\sin\theta \pa_\theta \phi)+{1\over \sin\theta} \phi
,_{\varphi \varphi} =0.
\end{eqnarray}

where $ \widetilde{\Delta}(E_\phi)\equiv\Theta E_\phi /2$, and the
following formula has also been noticed
\begin{equation}
\sum_{n=0}^\infty \left( {\widetilde{\Delta}( E_\phi)\over
r-r_H}\right)^{2n}=\left( 1-{\widetilde{\Delta}(E_\phi)^2\over
(r-r_H)^2} \right)^{-1}.
\end{equation}
The effect action in the range $\mathcal{R}_I $ is
\begin{eqnarray}
I_{eff}=-{1\over 2}\int d^4x
\sqrt{-\tu{g}}\tu{g}^{\mu\nu}(\partial_\mu\phi
\partial_\nu\phi)\;\;\;r\in\mathcal{R}_I.
\end{eqnarray}
The equation of motion of the $\phi$-field get from the above
action is
\begin{eqnarray}\label{effectem}
\pa_\mu(\sqrt{-\tu{g}}\tu{g}^{\mu\nu}\pa_\nu\phi)=0
\end{eqnarray}
Comparing the eq.(\ref{em0}) with the eq.(\ref{effectem}), we have
\begin{eqnarray}\label{eff}
&&\sin ^2 \theta \widetilde{g}_{\theta \theta}
=\widetilde{g}_{\varphi\varphi},\;\;\;
\widetilde{g}_{tt}\widetilde{g}_{rr} = -1 \\ \label{eff3}\nonumber
&&
\widetilde{g}_{tt}= ((r-r_-)(r-r_H)+\tu{\Delta}(E_\phi)^2
)^{1\over 2}( {r^4 \over (r-r_-)(r-r_H)}\\ \nonumber
&&\;\;\;\;\;+\tu{\Delta}(E_\phi)^{2}
 + {1 \over r_H-r_-}[ {r_H^4 \over r-r_H} \left({1\over
1-\left({\tu{\Delta}(E_\phi) \over r-r_H} \right)^2}-1\right)\\ &&
\;\;\;\;\;- {r_-^4 \over r-r_-} \left({1\over
1-\left({\tu{\Delta}(E_\phi) \over r-r_-} \right)^2}-1\right)
])^{-{1\over 2}}
   \\ \nonumber
\label{eff4}&& \widetilde{g}_{\theta \theta} =
\sqrt{(r-r_-)(r-r_H)+\tu{\Delta}(E_\phi)^2} ( {r^4 \over
(r-r_-)(r-r_H)}\\ \nonumber &&\;\;\;\;\;+\tu{\Delta}(E_\phi)^{2}
 + {1 \over r_H-r_-}[ {r_H^4 \over r-r_H} \left({1\over
1-\left({\tu{\Delta}(E_\phi) \over r-r_H} \right)^2}-1\right)\\ &&
- {r_-^4 \over r-r_-} \left({1\over 1-\left({\tu{\Delta}(E_\phi)
\over r-r_-} \right)^2}-1\right)  ])^{-{1\over 2}}
\end{eqnarray}
Solving the equation $\widetilde{g}_{tt}(r)=0$, we get two new
horizons with $r=r_H\pm \tu{\Delta}(E_\phi)$, and hence
$\Delta=\Delta _0(E_\phi)=\tu{\Delta}(E_\phi)$. Thus the quantum
horizon spread range has been determined.
In addition, note also that the $ \widetilde{g}_{\theta\theta} $
blows up at these two points which imply that the curvature scalar
vanishes there, too. Therefore they are regular. Note that, as
also discussed in \cite{thooft96}, the energy $E_\phi$ for the
scalar $\phi$ should not be too large, therefore
$\Delta(E_\phi)=\Theta E_\phi/2=l_p(l_pE_\phi/2)$ is not larger
than the Planck length $l_p$.

Using WKB and S-wave approximation, i.e., $\phi=\exp (-iE_\phi t
-i\int k(r)dr)$, and the equation of motion (\ref{NMT1}) and the
usual dispersion relation due to the commutative action
(\ref{Nac1}) and eq.(\ref{EMII}) , we obtain the wave number
$k(r)$ with given energy $E_\phi$ in the both noncommutative and
commutative ranges as follows
\begin{eqnarray}\label{cwn}\nonumber
&& \hskip-0.3in k^2(r)_{\mathcal{R}_I}=\left({1\over
x^2-1}\right){E_\phi^2 r_H^4\over \Delta^2(r_H-r_-)^2} {1\over
\left(x^2+{x\xi \over
1-a+x\xi}\right)\left(1+{x\xi \over 1-a}\right)}   \\
\nonumber
&&\;\;\;\;\;\;\;\times\{(x^2-1)[(1+x\xi)^4+\xi^4x(x+{1-a\over
\xi})]+1+{x\xi\over 1-a}\\&&\;\;\;\;\;\;\; - {a^4\over
1-a}x(x^2-1)\xi\left( {1\over
1-{ \xi^2 \over (x\xi +1-a)2}}-1\right)\} \\
\label{cwn1} && k^2(r)_{\mathcal{R}_{II}}={E_\phi^2 r^4 \over
(r-r_H)^2(r-r_-)^2}
\end{eqnarray}
where $x={(r-r_H) / \Delta}$,$\xi={\Delta / r_H}$,$a={r_-/
r_H}$.Denoting that, to the large hole $r_H>> \Delta$, $\xi\approx
0$, then we have
\begin{eqnarray}\label{dispersion2}
k(r)_{\mathcal{R}_I}&\simeq & \pm i{r_H^2E_\phi \over
\sqrt{1-x^2}(r_H-r_-)\Delta}.
\end{eqnarray}

Note the wave number $k$ in ${\cal R}_I$ is imaginary. This means
that the quantum tunnelling occurred in this noncommutative range.
The ratio between the amplitude of outgoing wave function of
$\phi$ outside the hole and one inside it is
$\exp({2i\int^{r_H-\Delta}_{r_H+\Delta} k(r)dr})$, which describes
the the quantum tunnelling effects of $\phi$ passing the
non-commutative range of the black hole. Using
eq.(\ref{dispersion2}), the following integration can be be
calculated
\begin{eqnarray}\label{nkk}
&& Im\left( 2\int k(r)dr\right)=
 \pm {2E_\phi r_H^2\over r_H-r_-}\int ^{+1}_{-1} {dx\over
\sqrt{1-x^2}}\\ \nonumber &&= \pm {2E_\phi r_H^2\over r_H-r_-}
\arcsin (x)|^1_{-1}=\pm {2\pi E_\phi r_H^2\over r_H-r_-}
\end{eqnarray}
 By means of the theory of
Dammur-Ruffini\cite{Ruffini} and Sannan\cite{Sannan}, the black
hole temperature $T_{BH}$ can be derived from the ratio between
the amplitude of outgoing wave function outside the hole and one
inside the hole, i.e., $\exp(|{2i\int^{r_H-\Delta}_{r_H+\Delta}
k(r)dr}|)=\exp (E_\phi /(2T_{BH}))$. Thus, using (\ref{nkk}), we
get the desired result as follows
\begin{equation}\label{T1}
T_{BH}={r_H-r_- \over 4r_H^2\pi}={\kappa\over 2\pi},
\end{equation}
where $\kappa={1\over 2}(r_H-r_-)/r_H^2$ is the surface gravity on
event horizon of the given metric (\ref{metric}). Then, the
standard Hawking temperature has been reproduced successfully.
 We like to address that our
 deriving of $T_{BH}$ is more natural than ordinary
 ones in the
 literatures\cite{Hawking74}\cite{Ruffini}\cite{Wilczek}. In order
to see this point, setting the non-commutative parameter
$\Theta=0$ and then $k(r)\rightarrow k_0(r)=E_\phi
r^2/((r-r_H)(r-r_-))$ (see eqs.(\ref{cwn})(\ref{cwn1})), our model
back to the ordinary commutative QFT in the curved space with
classical horizon used by ref.\cite{Ruffini}. In this case, the
calculations of imaginary part of the wave-number integral across
the horizon are as follows
\begin{eqnarray}\label{dr}
Im\left({2 E_\phi}\lim_{\epsilon
 \rightarrow 0}\int {r^2dr\over (r-r_-)(r-r_H \mp i\epsilon)}\right)=\pm {2\pi E_\phi r_H^2\over
r_H-r_-}.
\end{eqnarray}
Two remarks on the calculations of (\ref{dr}) are in order: 1, The
integration $\int dr k_0(r)$ is actually divergent due to the
singularity at $r=r_H$ in the integrand, and hence $Im\left( 2\int
k_0(r)dr\right)$ is somehow meaningless as without a proper
regularization; 2, The calculation (\ref{dr}) just serves as such
a regularization  which belongs to a (non-rigorous) mathematical
trick. This is, of course, {\it ad hoc}. Comparing eq.(\ref{nkk})
with eq.(\ref{dr}), we can see that the integration of (\ref{nkk})
is regular, and the mechanism to come over the ambiguity mentioned
in above is subtle. So that, we claim that the considerations to
quantum properties of the event horizon (i.e., quantum horizon)
presented in this letter have indeed solved the puzzle in the
Hawking radiation calculations existed in the literature caused by
the singularity at the classical horizon naturally.

  Because of the appearance of the new horizon, an observer cannot
detect the $\phi$ at $r\leq r_H +\Delta(E_\phi)$. Just we have
discussed earlier that the field has an infinite ret-shift at
$r=r_H +\Delta(E_\phi)$. We can now follow 't Hooft's
method\cite{thooft85} to evaluate the black hole entropy in our
model, but we emphasize again that there is no brick wall in our
model, and  $\Delta(E_\phi)$ is derived by the action (\ref{Nac}).
 For simplicity, we only study the singlet scale field in the above. It is
straightforward to extend the studies of actions of (\ref{Nac})
(\ref{Nac1}) of $\phi$ to ones of $\phi^A$ with $A=1\cdots N$
where $N$ is the number of the  fields. As well known that all the
fields couple with the gravity force which emitting from the
vacuum near the horizon, also, all of their dynamical degrees
freedom will contribute to entropy.
 For example, two-component
spin field contribution to the entropy is double of which of the
scale field. The WKB wave number $k(r)$ read from the action
eq.(\ref{Nac1}) is
\begin{eqnarray}
k^2(r)_{\mathcal{R}_{II}}={E_\phi^2r^4\over (r-r_-)^2
(r-r_H)^2}-{l(l+1)\over (r-r_H)(r-r_-)},
\end{eqnarray}
The number of states below energy $E_\phi$ is
\begin{equation}
g(E_\phi)=N \int dl(2l+1)\int^L_{r_H+\Delta(E_\phi)} dr
\sqrt{k^2(r)_{\mathcal{R}_{II}}}
\end{equation}
where $L$ represents an infrared cutoff. The free energy then
reads
\begin{equation}
\pi \beta F= \int dg( E_\phi) \ln \left(1-e^{-\beta E_\phi}
\right)= -\int^\infty_0 d E_\phi {\beta g(E_\phi)\over e^{\beta
E_\phi}-1}
\end{equation}
where $\beta =1/T_{BH}$ is inverse temperature. The dominant
contribution from the even horizon to $F$ is
\begin{equation}
F\approx - {8N\zeta (3) r_H^6\over 3\pi (r_H-r_-)^2l_p^2 \beta^3}
\end{equation}
where $\zeta(3)$ is Riemann $\zeta$-function,
$\zeta(3)=\sum^\infty_{n=1}1/n^3 \approx 1.202$. The entropy of
the black hole  now be obtained as
\begin{equation}\label{entropy}
S=\beta^2{\pa F\over \pa \beta}= {N\zeta(3)A\over 8\pi^4 l^2_p}
\end{equation}
where $A=4\pi r_H^2$ is the horizon area. So we reproduce the
correct relation of $S\propto A$. Consequently, combining it with
our Hawking radiation derivations presented in above (see
eqs.(\ref{cwn})--(\ref{T1})), we conclude that the black hole
entropy and the Hawking temperature have been produced
successfully and simultaneously in the QFT-model in curved space
with quantum horizon.
Thus we have achieved to bridge those two black hole physics
phenomena consistently by one semi-classical quantum field theory.
All puzzles caused by the singularity associating with the
classical horizon mentioned in the beginning of this letter have
been solved by considering the quantum effects on the horizon.

Finally, let us turn to discuss the entropy normalization. The
thermodynamics requires the black hole entropy has to be
normalized to the Bekenstein-Hawking expression, i.e.,
$S=S_{BH}=A/(4l_p^2)$ (see eq.(\ref{s1})). Then, from the
eq.(\ref{entropy}) we obtain the field number $N$  as follows
\begin{equation}\label{N}
N=2\pi^4/ \zeta(3)\approx 162,
\end{equation}
where $N$ fails to be a integer  because some approximations
(e.g., WKB-approximation etc) in the state-counting are used. Such
sort of errors
 should be rather small. Eq.(\ref{N}) is a good
 estimate to the $\phi$-field number
 in the model. Noting this field
 number can not be predicted in other types of brick wall models
 because there exist additional parameters.

It is well known that  the particle-antiparticle(with negative
energy)-pair creation effects in physical vacuum induced by the
strong classical gravitational fields will occur near the horizon
and it will lead to the Hawking radiations.
Because the gravitational interactions near the black hole horizon
are so  strong  that the interactions in the ordinary particle
dynamics can be ignored, the matter quantum dynamics near the
black hole will reduce  to action of free massless scalar fields
within the background gravitational fields like
eqs(\ref{Nac})(\ref{Nac1}),
but the field modes should keep to be unchangeable, and then the
number of degree of freedom of the fields are same as one in the
case without such kind of extremely strong background
gravitational fields, i.e., the original dynamical Lagrangian
case. Consequently, through counting the black hole's entropy and
normalizing it to be $S_{BH}$, we could get the field mode number
$N$ in the QFT model with quantum horizon, and then the number of
field degree of freedom in the underlying dynamics will be
determined. Here the interesting criterion is that those two
numbers must be equal each other in principle, because all
particle-pair created from the vacuum must belong to the
underlying matter theory and vice-versa, i.e., all particles in
the theory must be created by the extremely strong gravitational
interactions. Therefore, the mode number $N$ should be independent
of the parameters of black hole, i.e., $N$ displays an intrinsic
character of vacuum. That the black hole physics result could be
helpful for searching the new physics beyond the standard model is
remarkable.

Back to our result eq.(\ref{N}). Surely, $N\simeq 162$ is much
larger than the number of degree of freedom of the Standard Model
(SM) , which is 82 (see the TABLE). Therefore, at Planck scale, SM
fails. To the Minimal Super-symmetric Standard Model (MSSM), the
dynamical degree number is 164 (see TABLE) , which is very close
to our prediction $N=162$. Hence, our model favor to support MSSM
as a new physics theory beyond SM.

\begin{footnotesize}
\begin{tabular}{|l|l|c|l|l|c|}\hline
   \multicolumn{6}{|c|}{\bfseries Multiple number of fields in }\\
   \multicolumn{6}{|c|}{\bfseries minimal supersymmetric standard model\cite{Haber}} \\
\hline
Normal$\;\;\;\;\;\;\;\;$ & Name & N& Suppersymme- & Name & N \\
particles &    &   &tric partners & &  \\
\hline \hline $q=u,d,s,$ & quark & & $\tu{q}_L$, $\tu{q}_R$ &
scalar-quark&
\\ $\;\;\;\;\;\;\; c,b,t$ &$\;$($\times 3$ color) & 36 & &$\;\;\;\;\;\;$ ($\times 3$ color)&36 \\
\hline $l=e,\mu,\tau$ & lepton & 6 & $\tu{l}_L$, $\tu{l}_R$ &
scalar-lepton & 6 \\ \hline
 $\nu = \nu_e , \nu_\mu , \nu_\tau $ &  neutrino & 6 & $\tu{\nu}$
 & scalar-neutrino & 6 \\ \hline
 g & gluon & 16 & $\tu{\rm g}$ & gluino & 16 \\ \hline
 $W^{\pm}$ &  & 4 & $\tu{W}^\pm $ & wino & 4 \\ \hline
 $Z^0$ & & 2& $\tu{Z}^0$ & zino & 2 \\ \hline
 $\gamma$ & photon & 2 & $\tu{\gamma}$ & photino & 2 \\ \hline
 $H^+_1 \;\;\;\;\;\; H^0_1$ & & &$\tu{H}^+_1 \;\;\;\;\;\;
 \tu{H}^0_1$& & \\
$H^+_1 \;\;\;\;\;\; H^0_1$ & higgs & 8 &$\tu{H}^+_1 \;\;\;\;\;\;
\tu{H}^0_1$ & higgsino & 8 \\ \hline $g_{\mu\nu}$ & graviton & 2 &
$\tu{g}_{\mu\nu}$ & gravitino & 2 \\ \hline
\end{tabular}
\end{footnotesize}

The studies in this letter is in Schwarzchild-like coordinate, to
examine the conclusions in other coordinates (e.g.,
Painlev$\acute{e}$-, or Lemaitre codinates\cite{Jing} etc) remans
to be further study objects.

\begin{center}
{\bf ACKNOWLEDGMENTS}
\end{center}
 This work is partially supported by NSF of China 90403021 and
the PhD Program Fund of China. The authors wish to thank Dao-Neng
Gao,
 Jian-Xin Lu, and Shuang-Qing Wu  for their
stimulating discussions.


\begin{thebibliography}{99}
\bibitem{by2} In the ref.\cite{by03}, we set $E$ to be ADM-mass.It is not approperiate.
\end{thebibliography}
\end{document}